\documentclass[12pt]{article}
\usepackage{graphicx}
\def\hybrid{\topmargin 0pt      \oddsidemargin 0pt
        \headheight 0pt \headsep 0pt
        \voffset=-0.5cm
        \textwidth 6.25in       
        \textheight 9.5in       
        \marginparwidth 0.0in
        \parskip 5pt plus 1pt   \jot = 1.5ex}
\catcode`\@=11
\def\marginnote#1{}

\newcount\hour
\newcount\minute
\newtoks\amorpm
\hour=\time\divide\hour by60
\minute=\time{\multiply\hour by60 \global\advance\minute by-\hour}
\edef\standardtime{{\ifnum\hour<12 \global\amorpm={am}%
        \else\global\amorpm={pm}\advance\hour by-12 \fi
        \ifnum\hour=0 \hour=12 \fi
        \number\hour:\ifnum\minute<10 0\fi\number\minute\the\amorpm}}
\edef\militarytime{\number\hour:\ifnum\minute<10 0\fi\number\minute}

\def\draftlabel#1{{\@bsphack\if@filesw {\let\thepage\relax
   \xdef\@gtempa{\write\@auxout{\string
      \newlabel{#1}{{\@currentlabel}{\thepage}}}}}\@gtempa
   \if@nobreak \ifvmode\nobreak\fi\fi\fi\@esphack}
        \gdef\@eqnlabel{#1}}
\def\@eqnlabel{}
\def\@vacuum{}
\def\draftmarginnote#1{\marginpar{\raggedright\scriptsize\tt#1}}
\def\draftlabel#1{{\@bsphack\if@filesw {\let\thepage\relax
   \xdef\@gtempa{\write\@auxout{\string
      \newlabel{#1}{{\@currentlabel}{\thepage}}}}}\@gtempa
   \if@nobreak \ifvmode\nobreak\fi\fi\fi\@esphack}
        \gdef\@eqnlabel{#1}}
\def\@eqnlabel{}
\def\@vacuum{}
\def\draftmarginnote#1{\marginpar{\raggedright\scriptsize\tt#1}}

\def\draft{\oddsidemargin -.5truein
        \def\@oddfoot{\sl preliminary draft \hfil
        \rm\thepage\hfil\sl\today\quad\militarytime}
        \let\@evenfoot\@oddfoot \overfullrule 3pt
        \let\label=\draftlabel
        \let\marginnote=\draftmarginnote
   \def\@eqnnum{(\theequation)\rlap{\kern\marginparsep\tt\@eqnlabel}%
\global\let\@eqnlabel\@vacuum}  }

\def\numberbysection{\@addtoreset{equation}{section}
        \def\theequation{\thesection.\arabic{equation}}}

\def\underline#1{\relax\ifmmode\@@underline#1\else
        $\@@underline{\hbox{#1}}$\relax\fi}

\def\titlepage{\@restonecolfalse\if@twocolumn\@restonecoltrue\onecolumn
     \else \newpage \fi \thispagestyle{empty}\c@page\z@
        \def\thefootnote{\fnsymbol{footnote}} }

\def\endtitlepage{\if@restonecol\twocolumn \else  \fi
        \def\thefootnote{\arabic{footnote}}
        \setcounter{footnote}{0}}  
\relax

\numberbysection
\hybrid

\newfont{\Bbb}{msbm10 scaled 1\@ptsize00}

\newcommand{\ZZ}{\mbox{\Bbb Z}}
\newfont{\Bbbb}{msbm7 scaled 1\@ptsize00}
\newcommand{\cc}{\raise-1pt\hbox{$\mbox{\Bbbb C}$}}
\newcommand{\zz}{\raise-1pt\hbox{$\mbox{\Bbbb Z}$}}

\newtheorem{lemma-definition}{Lemma-Definition}[section]

\def\beq{\begin{equation}}
\def\eeq{\end{equation}}
\def\p{\partial}

\begin{document}

\begin{titlepage}

\title{Time discretization 
of the deformed Ruijsenaars-Schneider system}

\author{
A.~Zabrodin\thanks{
Skolkovo Institute of Science and Technology, Moscow 143026, 
Russia and
National Research University Higher School of Economics,
20 Myasnitskaya Ulitsa,
Moscow 101000, Russia and
NRC ``Kurchatov Institute'', Moscow, Russia;
e-mail: zabrodin@itep.ru}}

\date{January 2023}
\maketitle

\vspace{-7cm} \centerline{ \hfill ITEP-TH-04/23}\vspace{7cm}

\vspace{0.5cm}

\begin{abstract}

We obtain B\"acklund transformations and 
integrable time discretization of the recently introduced deformed
Ruijsenaars-Schneider many-body system which is the dynamical system
for poles of elliptic solutions to the Toda lattice with constraint
of type B. We also show that the deformed Ruijsenaars-Schneider system in discrete time is the dynamical system for poles of elliptic
solutions to the fully discrete Kadomtsev-Petviashvili equation
of type B. Besides, we suggest a field analogue of the deformed 
Ruijsenaars-Schneider system on a space-time lattice.

\end{abstract}


\end{titlepage}

\tableofcontents

\vspace{5mm}

\section{Introduction}

Integrable many-body systems of Calogero-Moser (CM) and
Ruijsenaars-Schneider (RS) type play a significant
role in modern mathematical physics. They are important and meaningful
from both mathematical and physical points of view (see the review
\cite{OP81}). 
An interesting problem of their integrable time discretization
was addressed in \cite{NP94,NRK96}, 
see also the book \cite{Suris}. This problem is closely connected
with B\"acklund transformations of the CM and RS systems
\cite{W82,BSTV14,ZZ18} and the so-called self-dual form of 
their equations of motion \cite{ABW09,Z19}. The idea is that 
the B\"acklund transformation can be regarded as evolution in
discrete time by one step.

The equations of motion of the most general elliptic version of the 
$N$-particle 
RS system in discrete time obtained
in \cite{NRK96} are as follows:
\beq\label{int1}
\begin{array}{l}
\displaystyle{
\prod_{k=1}^N \sigma (x_i^{n}-x_k^{n+1}-\eta )
\sigma (x_i^{n}-x_k^{n}+\eta )
\sigma (x_i^{n}-x_k^{n-1})}
\\ \\
\displaystyle{ \phantom{aaaaaaa}+\, \prod_{k=1}^N
\sigma (x_i^{n}-x_k^{n+1})
\sigma (x_i^{n}-x_k^{n}-\eta )\sigma (x_i^{n}-x_k^{n-1}+\eta )}=0,
\end{array}
\eeq
where $x_i^n$ is coordinate of the $i$th particle at the $n$th step
of discrete time and $\eta$ plays the role of a lattice spacing in the
space lattice. 
Here and below we use the 
standard Weierstrass $\sigma$-, $\zeta$- and $\wp$-functions 
$\sigma (x)$, $\zeta (x)=\sigma ' (x)/\sigma (x)$
and $\wp (x)=-\zeta '(x)$ (see Appendix A for their definition and 
properties).
The properly taken continuous time limit of (\ref{int1}) 
yields the equations of
motion of the RS model \cite{RS86,Ruij87}:
\beq\label{int2}
\ddot x_i +\sum_{j\neq i}^N \dot x_i \dot x_j
\Bigl (\zeta (x_{ij}+\eta )+
\zeta (x_{ij}-\eta )-2\zeta (x_{ij})\Bigr )=0, \quad x_{ij}\equiv
x_i-x_j, 
\eeq
where dot means the time derivative.

Recently, a deformation of the RS model was introduced \cite{KZ22}
as a dynamical system describing time evolution of poles of 
elliptic solutions
to the Toda lattice with the constraint of type B \cite{KZ22a}.
Equations of motion of the deformed RS system are
\beq\label{int3}
\ddot x_i +\sum_{j\neq i}^N \dot x_i \dot x_j
\Bigl (\zeta (x_{ij}+\eta )+
\zeta (x_{ij}-\eta )-2\zeta (x_{ij})\Bigr )+\sigma (2\eta )
(U_i^--U_i^+)=0,
\eeq
where
\beq\label{int4}
U_i^{\pm}=\prod_{j\neq i}^N 
\frac{\sigma (x_{ij}\pm 2\eta )
\sigma (x_{ij}\mp \eta )}{\sigma (x_{ij}\pm \eta )\sigma (x_{ij})}.
\eeq
They differ from (\ref{int2}) be the additional terms
$\sigma (2\eta )
(U_i^--U_i^+)$.

In \cite{KZ22,Z22} it was shown that the system (\ref{int3})
can be obtained by restriction of certain Hamiltonian flow of the 
$2N$-particle RS system to the $2N$-dimensional subspace
${\cal P}\subset {\cal F}$ of the $4N$-dimensional 
phase space ${\cal F}$ 
corresponding to the configurations in which the $2N$ particles
stick together joining in $N$ pairs such that the distance between
particles in each pair is equal to $\eta$. 
The restriction gives equations (\ref{int3}), with
the $x_i$ being 
coordinate of the $i$th pair moving as a whole thing 
with the fixed distance
between the two particles. Using this observation, all integrals 
of motion of the system (\ref{int3}) has been found in \cite{Z22}. 

In this paper we suggest an integrable time discretization of the 
deformed RS system (\ref{int3}) obtained in the way similar to the one
which leads to (\ref{int1}) from (\ref{int2}). The equations of motion
in discrete time are:
\beq\label{int5}
\begin{array}{l}
\displaystyle{
\mu \prod_{k=1}^N \sigma (x_i^{n}-x_k^{n+1})
\sigma (x_i^{n}-x_k^{n}+\eta )
\sigma (x_i^{n}-x_k^{n-1}-\eta )}
\\ \\
\displaystyle{ \phantom{aaaaaaa}+\, \mu \prod_{k=1}^N
\sigma (x_i^{n}-x_k^{n+1}+\eta )
\sigma (x_i^{n}-x_k^{n}-\eta )\sigma (x_i^{n}-x_k^{n-1})}
\\ \\
\displaystyle{=\, 
\mu^{-1} \prod_{k=1}^N \sigma (x_i^{n}-x_k^{n+1}-\eta )
\sigma (x_i^{n}-x_k^{n}+\eta )
\sigma (x_i^{n}-x_k^{n-1})}
\\ \\
\displaystyle{ \phantom{aaaaaaa}+\, \mu^{-1}\prod_{k=1}^N
\sigma (x_i^{n}-x_k^{n+1})
\sigma (x_i^{n}-x_k^{n}-\eta )\sigma (x_i^{n}-x_k^{n-1}+\eta )},
\end{array}
\eeq
where $\mu$ is a parameter related to the lattice spacing in the time
lattice. The structure of each side is the same as the structure
of the left hand side of (\ref{int1}). 
We note that this form of equations of motion was conjectured
by I.Krichever some time ago. There are two different continuous
time limits of equations (\ref{int5}). One of them gives equations
of motion of the deformed RS system (\ref{int3}). The other one 
leads to the equations obtained in \cite{RZ20} as equations of motion
for poles of elliptic solutions to the semi-discrete Kadomtsev-Petviashvili
equation of type B (BKP).

Like the equations of motion for the RS system, equations (\ref{int5})
admit trigonometric (or hyperbolic) and rational degenerations. 
In the trigonometric limit, one of the two quasi-periods of the 
$\sigma$-function tends to $\infty$ and $\sigma (x)\to \sin x$.
In the rational limit, both quasi-periods tend to $\infty$ and 
$\sigma (x)\to x$. In this paper 
we will not discuss the specifics and details of the 
trigonometric and rational limits. 

We also present the extension of the deformed RS model to a
lattice field theory (a ``field analogue'') in which the
coordinates of particles $x_i$ become ``fields'' $x_i(x,t)$ 
depending not only
on the time $t$ but also on a space variables $x$. 
Following the method developed in \cite{AKV02} for the CM model
and applied in \cite{ZZ21} to the RS model, 
we obtain the equations of motion 
as equations
for poles of more general elliptic solutions (called 
elliptic families in \cite{AKV02}) to the fully discrete BKP equation.

The paper is organized as follows. In Section 2 we obtain the 
B\"acklund transformation of the deformed RS system. In Section 3 
the chain of B\"acklund transformations is interpreted as a discrete
time evolution and the equations of motion in discrete time are 
obtained. Possible continuum limits are also discussed. In Section 4
we show that
the discrete time equations of motion for 
the deformed RS system (\ref{int5}) 
describe the dynamics of poles of elliptic
solutions to the fully discrete BKP equation. Section 5 is devoted to
obtaining the lattice 
field analogue of the fully discrete deformed RS system.
Concluding remarks are
presented in Section 6. In Appendix A the definition and main 
properties of the Weierstrass functions are given. 

\section{The B\"acklund transformation}

Since the works \cite{AMM77}--\cite{KZ95} it
became a common knowledge that the integrable 
many-body systems of CM and RS type 
are dynamical systems for poles of singular solutions 
to nonlinear integrable differential and difference equations. 
The nonlinear integrable equations are known to serve as compatibility
conditions for linear differential or difference equations for 
the ``wave function'' $\psi$. Poles of solutions to the nonlinear
equations (zeros of the tau-function) are simultaneously poles 
of the $\psi$-function, so the latter are subject to equations of
motion of the CM or RS type. In fact zeros of the $\psi$-function 
are subject to the same equations, and this leads to the idea 
to obtain the B\"acklund transformation of the CM or RS system as 
passage from poles to zeros. This approach works for all previously 
known examples, and we are going to apply it to the case of the 
Toda lattice with constraint of type B.

The first linear problem for the Toda lattice
with constraint of type B has the form \cite{KZ22a}
\beq\label{b1}
\p_t \psi (x) =v(x)\Bigl (\psi (x+\eta )-\psi (x-\eta )\Bigr ),
\eeq
where $v(x)$ is expressed through the tau-function $\tau (x)$ as
\beq\label{b2}
v(x)=\frac{\tau (x+\eta )\tau (x-\eta )}{\tau^2(x)}.
\eeq
For elliptic solutions
\beq\label{b3}
\tau (x)=\prod_{i=1}^N \sigma (x-x_i),
\eeq
where the $x_j$'s are assumed to be all distinct,
so $v(x)$ is an elliptic function with periods $2\omega$, $2\omega '$. 
Therefore, solutions to (\ref{b1}) can be sought as double-Bloch 
functions, i.e. functions such that
$$
\psi (x+2\omega )=B\psi (x), \quad \psi (x+2\omega ')=B'\psi (x)
$$
with some Bloch multipliers $B$, $B'$. Poles of the $\psi$-function are
zeros of the tau-function. Therefore, we can represent solutions
of (\ref{b1}) in the form
\beq\label{b4}
\psi (x)=\mu^{x/\eta}e^{(\mu -\mu^{-1})t} \, 
\frac{\hat \tau (x)}{\tau (x)},
\eeq
where
\beq\label{b5}
\hat \tau (x)=\prod_{i=1}^N \sigma (x-y_i)
\eeq
with some $y_i$'s.
Then the $\psi$-function is indeed a double-Bloch function with
the Bloch multipliers
$$
B=\mu^{2\omega /\eta}\exp \Bigl (2\zeta (\omega )
\sum_{j=1}^N (x_j-y_j)\Bigr ),
\quad
B'=\mu^{2\omega ' /\eta}\exp \Bigl (2\zeta (\omega ')
\sum_{j=1}^N (x_j-y_j)\Bigr ).
$$
Below we will see that 
$$
\sum_{j=1}^N (\dot x_j-\dot y_j)=0,
$$
so the Bloch multipliers do not depend on time. It is proved in 
\cite{KZ22} that poles of the $\psi$-function $x_j$ satisfy the equations
of motion (\ref{int3}) for any $\mu$.

Substituting the ansatz (\ref{b4}) into equation (\ref{b1}) with
$v(x)$ given by (\ref{b2}), we obtain the equation
\beq\label{b6} 
\frac{\p_t \hat \tau (x)}{\hat \tau (x)}-
\frac{\p_t \tau (x)}{\tau (x)}+\mu -\mu^{-1}=
\mu \frac{\hat \tau (x+\eta )\tau (x-\eta )}{\hat \tau (x)\, \tau (x)}-
\mu^{-1} \frac{\tau (x+\eta )\hat \tau (x-\eta )}{\tau (x)\, \hat \tau (x)}.
\eeq
It is evident from this equation that it is invariant under the
simultaneous changes $\tau \leftrightarrow 
\hat \tau$, $\mu \leftrightarrow \mu^{-1}$, so the $y_j$'s satisfy the
same equations of motion (\ref{int3}). Both sides of equation (\ref{b6})
have simple poles at $x=x_j$ and $x=y_j$. Equating the residues, we 
obtain the equations
\beq\label{b7}
\begin{array}{l}
\displaystyle{
\dot x_i =\mu \sigma (-\eta )\prod_{j\neq i}
\frac{\sigma (x_i\! -\! x_j\! -\! \eta )}{\sigma (x_i\! -\! x_j)}
\prod_{k}
\frac{\sigma (x_i\! -\! y_k\! +\! \eta )}{\sigma (x_i\! -\! y_k)}}
\\ \\
\displaystyle{\phantom{aaaaaaaaaaaaaaa}+
\mu^{-1} \sigma (-\eta )\prod_{j\neq i}
\frac{\sigma (x_i\! -\! x_j\! +\! \eta )}{\sigma (x_i\! -\! x_j)}
\prod_{k}
\frac{\sigma (x_i\! -\! y_k\! -\! \eta )}{\sigma (x_i\! -\! y_k)},}
\\ \\
\displaystyle{
\dot y_i =\mu \sigma (-\eta )\prod_{j\neq i}
\frac{\sigma (y_i\! -\! y_j\! +\! \eta )}{\sigma (y_i\! -\! y_j)}
\prod_{k}
\frac{\sigma (y_i\! -\! x_k\! -\! \eta )}{\sigma (y_i\! -\! x_k)}}
\\ \\
\displaystyle{\phantom{aaaaaaaaaaaaaaa}+
\mu^{-1} \sigma (-\eta )\prod_{j\neq i}
\frac{\sigma (y_i\! -\! y_j\! -\! \eta )}{\sigma (y_i\! -\! y_j)}
\prod_{k}
\frac{\sigma (y_i\! -\! x_k\! +\! \eta )}{\sigma (y_i\! -\! x_k)}.}
\end{array}
\eeq
These are so-called self-dual equations of motion. They are symmetric
under the exchange $x_j \leftrightarrow y_j$, 
$\mu \leftrightarrow \mu^{-1}$. The passage $x_j \to y_j$ can be regarded
as a B\"acklund transformation of the deformed RS system.
The equations of motion (\ref{int5}) for $x_j$'s and $y_j$'s 
in principal can be derived from (\ref{b7}). To this end, one should
take the time derivative of these equations and use them again 
substituting 
the expressions for $\dot x_j$, $\dot y_j$ through $x_j$, $y_j$.
Equations (\ref{int5}) are then equivalent to a non-trivial identity
for multi-variable elliptic functions which is too complicated to be
proved directly. However, we do not need to prove it directly 
since the equations
of motion for
the $x_j$'s follow from the result of
\cite{KZ22} and those for the $y_j$'s follow from the symmetry
$x_j \leftrightarrow y_j$. Note that the B\"acklund transformation
for the RS system differs from (\ref{b7}) by absence of the second
terms in the right hand sides. In this sense it is contained 
in (\ref{b7}) as a formal limiting case $\mu \to \infty$ (or 
$\mu \to 0$). 

From equations (\ref{b7}) it follows that
\beq\label{b8}
\sum_{j=1}^N (\dot x_j -\dot y_j)=0.
\eeq
The proof is based on the higher degree 
identities for the $\sigma$-function 
of the Fay type (see, e.g., \cite{ZZ18}). It is
almost the same as the corresponding proof from
\cite{ZZ18} and we omit it here. 

\section{The discrete time dynamics and continuum limits}

The B\"acklund transformation $x_j \to y_j$ can be regarded as 
a time evolution by one step of the discrete time. Denoting the 
discrete time variable by $n\in \ZZ$, we then write $x_i =x_i^n$,
$y_i=x_i^{n+1}$. Shifting $n\to n-1$ in the second equation in
(\ref{b7}), so that the left hand sides of the two equations 
become equal,
we conclude that the right hand sides are equal, too, which 
results in equations (\ref{int5}), or
\beq\label{d1}
\begin{array}{c}
\displaystyle{
\prod_{j=1}^N \frac{\sigma (x_i^{n}-x_j^{n+1})
\sigma (x_i^{n}-x_j^{n}+\eta )
\sigma (x_i^{n}-x_j^{n-1}-\eta )}{\sigma (x_i^{n}-x_j^{n+1}+\eta )
\sigma (x_i^{n}-x_j^{n}-\eta )\sigma (x_i^{n}-x_j^{n-1})}=-1}
\\ \\
\displaystyle{
+\, \mu^{-2}\prod_{j=1}^N \frac{\sigma (x_i^{n}\! -\! x_j^{n+1})
\sigma (x_i^{n}\! -\! x_j^{n-1}\! +\! \eta )}{\sigma (x_i^{n}
\! -\! x_j^{n+1}\! +\! \eta )
\sigma (x_i^{n}\! -\! x_j^{n-1})}+
\mu^{-2}\prod_{j=1}^N \frac{\sigma (x_i^{n}\! -\! x_j^{n+1}\! -\! \eta )
\sigma (x_i^{n}\! -\! x_j^{n}\! +\! \eta )}{\sigma (x_i^{n}
\! -\! x_j^{n+1}\! +\! \eta )
\sigma (x_i^{n}\! -\! x_j^{n}\! -\! \eta)}.}
\end{array}
\eeq 

Let us discuss continuous time limit of equations (\ref{d1}). 
In fact they admit different continuum limits. For one of them,
we introduce the variables
\beq\label{d2}
X_j^n =x_j^n -n\eta 
\eeq
and assume that these variables behave smoothly when the time
changes, i.e., $X_j^{n+1}=X_j^n +O(\varepsilon )$ as
$\varepsilon \to 0$, where
we introduce the lattice spacing $\varepsilon$ in the time lattice,
so that the continuous time variable is $t=n\varepsilon$. 
In terms of the variables $X_j^n$, equations (\ref{d1}) acquire
the form
\beq\label{d3}
\begin{array}{c}
\displaystyle{
\prod_{j=1}^N \frac{\sigma (X_i^{n}-X_j^{n+1}-\eta )
\sigma (X_i^{n}-X_j^{n}+\eta )
\sigma (X_i^{n}-X_j^{n-1})}{\sigma (X_i^{n}-X_j^{n+1})
\sigma (X_i^{n}-X_j^{n}-\eta )\sigma (X_i^{n}-X_j^{n-1}+\eta )}=-1}
\\ \\
\displaystyle{
+\, \mu^{-2}\! \prod_{j=1}^N \frac{\sigma (X_i^{n}\! -\! 
X_j^{n+1}\! -\! \eta )
\sigma (X_i^{n}\! -\! X_j^{n-1}\! +\! 2 \eta )}{\sigma (X_i^{n}
\! -\! X_j^{n+1})
\sigma (X_i^{n}\! -\! X_j^{n-1}\! +\! \eta )}}
\\ \\
\displaystyle{\phantom{aaaaaaaaa}+
\mu^{-2}\! \prod_{j=1}^N \frac{\sigma (X_i^{n}\! -\! X_j^{n+1}\! - \!2 \eta )
\sigma (X_i^{n}\! -\! X_j^{n}\! +\! \eta )}{\sigma (X_i^{n}
\! -\! X_j^{n+1})
\sigma (X_i^{n}\! -\! X_j^{n}\! -\! \eta)}.}
\end{array}
\eeq 
We should expand these equations in powers of $\varepsilon$ 
taking into account that
$$X_j^{n\pm 1}=X_j \pm \varepsilon \dot X_j 
+\frac{1}{2}\, \varepsilon^2 \ddot X_j +O(\varepsilon^3)$$
as $\varepsilon \to 0$. The expansion procedure is straightforward.
It is enough to expand up to the order $\varepsilon$. For consistency
of the expansion procedure one should require that $\mu^{-1}$ is of
order $\varepsilon$. Putting $\mu^{-1}=\varepsilon$, one obtains 
equations (\ref{int3}) for the $X_j$'s 
in the leading order $\varepsilon$. 

Another possibility is to assume that the original variables
$x_j^n$ are smooth when the time
changes, i.e.,
$$x_j^{n\pm 1}=x_j \pm \varepsilon \dot x_j 
+\frac{1}{2}\, \varepsilon^2 \ddot x_j 
+O(\varepsilon^3).$$
In this case one should expand equations (\ref{d1}). It is easy to
see that in general position, i.e. if $\mu^{-2}-1 =O(1)$ 
as $\varepsilon \to 0$, the leading order is $\varepsilon$ and the expansion 
gives the RS equations (\ref{int2}). However, if 
$\mu^{-2} -1 =O(\varepsilon )$, say, 
$\mu^{-2} =1+\alpha \varepsilon +O(\varepsilon^2)$, then the first order
gives the identity $0=0$ and one should expand up to the second
order in $\varepsilon$. The procedure is rather cumbersome but
straightforward. As a result, one obtains the equations derived in
\cite{RZ20} for dynamics of poles of elliptic solutions to the
semi-discrete BKP equation:
\beq\label{d4}
\begin{array}{l}
\displaystyle{
\sum_{j\neq i} (\ddot x_i \dot x_j -\dot x_i \ddot x_j)
\Bigl ( \frac{\wp '(\eta )}{\wp (x_{ij})-\wp (\eta )}-2\zeta (\eta )
\Bigr )}
\\ \\
\displaystyle{\phantom{aaaaaaaaaaaaa}
+\, \sum_k \sum_{j\neq i,k}\dot x_i \dot x_j \dot x_k
\, \frac{\wp '(x_{ij})}{\wp (x_{ij})-\wp (\eta )}\, \Bigl (
\frac{\wp '(\eta )}{\wp (x_{ij})-\wp (\eta )}-2\zeta (\eta )
\Bigr )}
\\ \\
\displaystyle{\phantom{aaaaaaaaaaaaaaaaaaaaaa}
-\, 2\zeta (\eta )\sum_{j\neq i} \dot x_i \dot x_j^2\,
\frac{\wp '(x_{ij})}{\wp (x_{ij})-\wp (\eta )}}
\\ \\
\displaystyle{\phantom{aaaaaaaaaaaaa}
-\alpha \Bigl (\ddot x_i +\sum_{j\neq i}\dot x_i \dot x_j
\Bigl (\zeta (x_{ij}+\eta )+\zeta (x_{ij}-\eta )-2\zeta (x_{ij})\Bigr )
\Bigr )=0}
\end{array}
\eeq
(here we correct the misprint in \cite{RZ20}). In contrast to
the previously discussed equations of motion, these equations are
not resolved with respect to the $\ddot x_j$'s. 

\section{The discrete time deformed RS system 
from fully discrete BKP equation}

In this section we show that the discrete time elliptic 
deformed RS system (\ref{int5}) 
is a dynamical system for poles of elliptic solutions to the 
fully discrete BKP equation \cite{Miwa82}.

We begin with the continuous BKP hierarchy \cite{DJKM83,DJKM82}. 
Let ${\bf t}=\{ t_1, t_3, t_5, \ldots \}$
be an infinite set of continuous ``times'' indexed by odd numbers. 
They are independent variables of the 
hierarchy. The dependent variable is the tau-function 
$\tau = \tau ({\bf t})$. The 
BKP hierarchy is encoded in the generating 
basic bilinear relation for the 
tau-function \cite{DJKM82}:
\beq\label{disc1}
\oint_{C_{\infty}}\frac{dz}{2\pi i z}\,
e^{\xi ({\bf t}, z)-\xi ({\bf t}', z)}
\tau \Bigl ({\bf t}-2[z^{-1}]\Bigr )\, \tau \Bigl ({\bf t}' +2[z^{-1}]\Bigr )=
\tau ({\bf t})\tau ({\bf t}')
\eeq
valid for all ${\bf t}$, ${\bf t}'$. Here we use the standard notation
$$
\xi ({\bf t}, z)=\sum_{k=1,3,5,\ldots}\!\! t_k z^k, \quad
{\bf t}\pm 2[z^{-1}]=\left \{ t_1 \pm \frac{2}{z}, t_3 \pm \frac{2}{z^3},
t_5 \pm \frac{2}{z^5}, \ldots \right \}.
$$
The contour $C_{\infty}$ is a big circle around infinity such that
the singularities coming from the factors with the tau-functions are
inside it and those coming from the exponential factor are outside it.

The discrete BKP equation is obtained as follows.
Put
\beq\label{disc2}
\tau ({\ell },m,n):=\tau \Bigl ({\bf t}-2{\ell }
[a^{-1}]-2m[b^{-1}]-2n[c^{-1}]\Bigr ).
\eeq
Setting $t_k'=t_k-2a^{-k}/k-2b^{-k}/k -2c^{-k}/k$ in
the bilinear relation, one can calculate the integral 
in the left hand side of (\ref{disc1}) with the help of the
residue calculus. (In doing so one 
should take into account that the poles at
the points $a,b,c$ which arise from
the exponential factor are outside the contour, 
and the contour should be shrunk to
infinity.)
As a result, one obtains from (\ref{disc1}) 
the fully discrete BKP equation for
$\tau ({\ell },m,n)$ first appeared 
in \cite{Miwa82}:
\beq\label{disc3}
\begin{array}{c}
(a+b)(a+c)(b-c)\tau ({\ell }+1)\tau (m+1, n+1)
\\ \\
\phantom{aaaaaaa}-(a+b)(b+c)(a-c)\tau (m+1)\tau ({\ell }+1, n+1)
\\ \\
\phantom{aaaaaaaaaaaaaa}+(a+c)(b+c)(a-b)\tau (n+1)\tau ({\ell }+1, m+1)
\\ \\
\phantom{aaaaaaaaaaaaaaaaaaaaa}=(a-b)(a-c)(b-c)\tau \, 
\tau ({\ell }+1, m+1, n+1).
\end{array}
\eeq
For simplicity of the notation,
we explicitly indicate only the variables that are subject to shifts.

Let us introduce the 
wave function $\psi ({\ell },m;z)$ by the formula
\beq\label{disc5}
\psi ({\ell },m;z)=e^{\xi ({\bf t},z)}
\left (\frac{a-z}{a+z}\right )^{{\ell }}
\left (\frac{b-z}{b+z}\right )^{m}
\frac{
\tau \Bigl ({\bf t}-2{\ell }[a^{-1}]-2m[b^{-1}]-2[z^{-1}]
\Bigr )}{\tau \Bigl ({\bf t}-2l[a^{-1}]-2m[b^{-1}]\Bigr )}.
\eeq
A direct calculation shows that the fully 
discrete BKP equation (\ref{disc3}) is equivalent to 
the following
linear equation for the wave function:
\beq\label{disc6}
\psi ({\ell },m+1) -\psi ({\ell }+1,m)=
\frac{a-b}{a+b}\, u({\ell },m) \Bigl (\psi ({\ell }+1, m+1)-
\psi ({\ell },m) \Bigr ),
\eeq
where
\beq\label{disc7}
u({\ell },m)=\frac{\tau \, 
\tau ({\ell}+1, m+1)}{\tau ({\ell}+1)\tau (m+1)}.
\eeq

So far all the discrete variables entered symmetrically. Now we are
going to break the symmetry by distinguishing one of the variables, 
say $\ell$, treating it as a space variable on the space lattice
with lattice spacing $\eta$. Accordingly, we
introduce the continuous variable
$x={\ell}\eta$ with $\ell \pm 1$ corresponding to
$x\pm \eta$. The discrete time variable is
$m$. It is convenient to change the notation as
$$
\tau ({\ell},m,n)\to \tau^{m} (x), \quad
\psi({\ell},m)\to \psi^{m}(x)
$$
(the variable $n$ is supposed to be fixed).
With this notation, equation (\ref{disc6}) acquires the form
\beq\label{disc8}
\psi^{m+1}(x)-\psi^m(x+\eta )=\kappa \, u^m(x)\Bigl (
\psi^{m+1}(x+\eta )-\psi^m(x)\Bigr ),
\eeq
\beq\label{disc9}
u^m(x)=\frac{\tau^m(x)\, \tau^{m+1}(x+\eta )}{\tau^{m}(x+\eta )\,
\tau^{m+1}(x)},  \quad \kappa = \frac{a-b}{a+b}.
\eeq

Let us point out how the two continuous time limits discussed in
the previous section look like in these terms.
For the first limit (which leads to the Toda lattice with constraint
of type B) one should tend $b\to a$ and $m\to \infty$ 
in such a way that the continuous
time variable $t=m(b^{-1}-a^{-1})$ remains finite and nonzero. 
For the second limit (which leads to
the semi-discrete BKP equation) one should tend $b\to \infty$ 
and $m\to \infty$ 
in such a way that the continuous time variable 
$t=mb^{-1}$ remains finite and nonzero.

Now we are going to obtain equations of motion 
for poles in $x$ of the wave function
$\psi^m (x)$ (which are zeros of the $\tau^m(x)$) as functions of 
the discrete
time $m$. According to (\ref{disc5}), we represent the wave function
in the form
\beq\label{disc10}
\psi^m(x)=\lambda^{x/\eta}\nu^m \, \frac{\hat \tau^m(x)}{\tau^m(x)}
\eeq
with
$\displaystyle{\lambda =\frac{a-z}{a+z}}$, 
$\displaystyle{\nu =\frac{b-z}{b+z}}$. Note that in terms of $\lambda$
and $\nu$ we have
\beq\label{disc11}
\kappa =\frac{\lambda -\nu}{1-\lambda \nu}.
\eeq
For elliptic solutions we set
\beq\label{disc12}
\tau^m(x)=\prod_{j=1}^N \sigma (x-x_j^m), \quad
\hat \tau^m(x)=\prod_{j=1}^N \sigma (x-y_j^m)
\eeq
and assume that the $x_j^m$'s are all distinct.
Plugging (\ref{disc10}) into the linear equation (\ref{disc8}), we 
obtain the following equation connecting $\tau^m(x)$ and $\hat \tau^m(x)$:
\beq\label{disc13}
\nu \frac{\hat \tau^{m+1}(x)}{\tau^{m+1}(x)}-
\lambda \frac{\hat \tau^{m}(x+\eta )}{\tau^{m}(x+\eta )}=
\lambda \nu \kappa\, \frac{\tau^m(x)
\hat \tau^{m+1}(x+\eta )}{\tau^{m}(x+\eta )\tau^{m+1}(x)}-
\kappa \, \frac{\hat \tau^m(x)
\tau^{m+1}(x+\eta )}{\tau^{m}(x+\eta )\tau^{m+1}(x)}.
\eeq
The both sides have simple poles at $x=x_i^{m+1}$ and $x=x_i^m-\eta$.
Equating the residues at these poles 
and making the shift
$m\to m-1$ when necessary, we get the equations
\beq\label{disc14}
\nu \tau^{m-1}(x_i^m \! +\! \eta )\hat \tau^m(x_i^m)=
\lambda \nu \kappa \tau^{m-1}(x_i^m)\hat \tau^m(x_i^m\! +\! \eta )-
\kappa \tau^m(x_i^m\! +\! \eta )\hat \tau^{m-1}(x_i^m),
\eeq
\beq\label{disc14a}
\lambda \tau^{m+1}(x_i^m \! -\! \eta )\hat \tau^m(x_i^m)=
\kappa \tau^{m+1}(x_i^m)\hat \tau^m(x_i^m\! -\! \eta )-
\lambda \nu \kappa \tau^m(x_i^m\! -\! \eta )\hat \tau^{m+1}(x_i^m).
\eeq
Substituting $x=x_i^m$ and $x=x_i^{m+1}\! -\! \eta$ into (\ref{disc13}),
we see that one term vanishes. 
Making the shift
$m\to m-1$ when necessary,
we get the equations
\beq\label{disc15}
\kappa \tau^{m+1}(x_i^m \! +\! \eta )\hat \tau^m(x_i^m)=
\lambda \tau^{m+1}(x_i^m)\hat \tau^m(x_i^m\! +\! \eta )-
\nu \tau^m(x_i^m\! +\! \eta )\hat \tau^{m+1}(x_i^m),
\eeq
\beq\label{disc15a}
\lambda \nu \kappa \tau^{m-1}(x_i^m \! -\! \eta )\hat \tau^m(x_i^m)=
\nu \tau^{m-1}(x_i^m)\hat \tau^m(x_i^m\! -\! \eta )-
\lambda \tau^m(x_i^m\! -\! \eta )\hat \tau^{m-1}(x_i^m).
\eeq
Equations (\ref{disc14}) and (\ref{disc15}) can be regarded as a
system of linear
equations for $\hat \tau^m(x_i^m)$ and $\hat \tau^m(x_i^m\! +\! \eta )$.
According to the Cramer's rule, the solution for $\hat \tau^m(x_i^m)$ is
\beq\label{disc16}
\hat \tau^m(x_i^m)=-\kappa \tau^m(x_i^m +\eta )\,
\frac{\left |\begin{array}{cc}
\hat \tau^{m-1}(x_i^m) & \lambda \nu \tau^{m-1}(x_i^m)
\\ & \\
\nu \hat \tau^{m+1}(x_i^m) & \lambda  \tau^{m+1}(x_i^m)
\end{array}\right |}{\left |\begin{array}{cc}
\nu \tau^{m-1}(x_i^m\! +\! \eta ) & -\lambda \nu \kappa \tau^{m-1}(x_i^m)
\\ & \\
-\kappa \tau^{m+1}(x_i^m\! +\! \eta ) & \lambda \tau^{m+1}(x_i^m)
\end{array}\right |}.
\eeq
In their turn,
equations (\ref{disc14a}) and (\ref{disc15a}) can be regarded as a
system of linear
equations for $\hat \tau^m(x_i^m)$ and $\hat \tau^m(x_i^m\! -\! \eta )$.
The solution for $\hat \tau^m(x_i^m)$ is
\beq\label{disc16a}
\hat \tau^m(x_i^m)=-\kappa \tau^m(x_i^m -\eta )\,
\frac{\left |\begin{array}{cc}
\lambda \nu \hat \tau^{m+1}(x_i^m) & \tau^{m+1}(x_i^m)
\\ & \\
\lambda \hat \tau^{m-1}(x_i^m) & \nu \tau^{m-1}(x_i^m)
\end{array}\right |}{\left |\begin{array}{cc}
-\lambda \tau^{m+1}(x_i^m\! -\! \eta ) & \kappa \tau^{m+1}(x_i^m)
\\ & \\
\lambda \nu \kappa \tau^{m-1}(x_i^m\! -\! \eta ) & -\nu \tau^{m-1}(x_i^m)
\end{array}\right |}.
\eeq
Equating the right hand sides of (\ref{disc16}) and (\ref{disc16a}),
we obtain the equations for the discrete time dynamics of poles $x_i^m$:
\beq\label{disc17}
\begin{array}{l}
\tau^{m+1}(x_i^m)\tau^m(x_i^m -\eta )\tau^{m-1}(x_i^m+\eta )+
\tau^{m+1}(x_i^m-\eta )\tau^m(x_i^m +\eta )\tau^{m-1}(x_i^m)
\\ \\
\phantom{}
=\, \kappa^2 \tau^{m+1}(x_i^m\! +\! \eta )\tau^m(x_i^m \! -\! 
\eta )\tau^{m-1}(x_i^m)
+\kappa^2 \tau^{m+1}(x_i^m)\tau^m(x_i^m \! +\! \eta )\tau^{m-1}(x_i^m
\! -\! \eta ).
\end{array}
\eeq
These are equations (\ref{int5}). 

\section{A field analogue of the deformed RS system on a space-time
lattice}

It is known that integrable models of the CM and RS type admit 
extensions to field theories (``field analogues'') in which the
coordinates of particles $x_i$ become ``fields'' $x_i(x,t)$ 
depending not only
on the time $t$ but also on a space variable $x$. Equations of motion
of these more general models can be obtained as equations
for poles of more general elliptic solutions (called 
elliptic families in \cite{AKV02}) to nonlinear integrable equations.
In this case, one considers solutions which are elliptic functions of a 
linear combination $\lambda$ 
of higher times $t_k$ of the hierarchy, their poles $\lambda_i(x,t)$
being functions of the space and time variables $x,t$ (in the 
CM/KP case $x=t_1$, $t=t_2$). They obey a system of partial differential
or difference equations which are equations of motion of the field
analogue of the CM or RS system. They were obtained by this method in \cite{AKV02} and \cite{ZZ21}
respectively (see also \cite{Z22a}, where 
elliptic families of solutions to the constrained Toda lattice were
discussed). 

In this section we apply this method to the fully discrete BKP
equation and obtain the field extension of the deformed RS model
on a space-time lattice. To wit, we will consider elliptic families 
of solutions to the fully discrete BKP equation and find dynamical
equations for their poles.

\subsection{Equations of motion on the space-time lattice from 
elliptic families for the fully discrete BKP equation} 

We begin with the BKP hierarchy (\ref{disc1}). Let 
$\displaystyle{\lambda =\sum_{j \; {\rm odd}}
\beta_j t_j}$ be an arbitrary linear
combination of the times of the hierarchy. According to \cite{AKV02},
the tau-function $\tau (\lambda , {\bf t})$ corresponding to a solution
which is an elliptic function of $\lambda$ is of the general form
\beq\label{f1}
\tau (\lambda , {\bf t})=\rho ({\bf t}) e^{c_1\lambda +c_2\lambda^2}
\prod_{j=1}^N \sigma (\lambda - \lambda_j ({\bf t})),
\eeq
where $\rho ({\bf t})$ does not depend on $\lambda$ and $c_1, c_2$
are some constants. Note that the shift of $\lambda$ by any period
should give an equivalent tau-function, i.e. the tau-function which 
differs from the initial one by multiplying by exponential function 
of a linear
form in the times. Therefore, the zeros $\lambda_i$
of the tau-function (poles of the solution) should satisfy the condition
\beq\label{f2}
\sum_i \lambda_i ({\bf t}) =\, \mbox{linear form in ${\bf t}$}.
\eeq
This means that 
the ``center of masses'' of the set of the $\lambda_i$'s
moves uniformly in all times. 

For the fully discrete BKP equation we can consider elliptic solutions
of the form
\beq\label{f3}
\tau^m (\lambda , x)=\rho^m (x) e^{c_1\lambda +c_2\lambda^2}
\prod_{j=1}^N \sigma (\lambda - \lambda_j^m (x)),
\eeq
where $x$ is a space variable and $m$ is the discrete time variable.
We assume that all $\lambda_j$'s are distinct.
Set
\beq\label{f4}
\sigma^m (\lambda , x):=
\prod_{j=1}^N \sigma (\lambda - \lambda_j^m (x)).
\eeq
We can find solutions to equation (\ref{disc8}) in the form
\beq\label{f5}
\psi^m(x)=\frac{\hat \tau^m(\lambda , x)}{\sigma^m(\lambda , x)}.
\eeq
Plugging (\ref{f3}), (\ref{f5}) into (\ref{disc8}), we get the equation
\beq\label{f6}
\begin{array}{c}
\displaystyle{
\frac{\hat \tau^{m+1}(\lambda , x)}{\sigma^{m+1}(\lambda , x)}\, -\, 
\frac{\hat \tau^{m}(\lambda , x+\eta )}{\sigma^{m}(\lambda , x+\eta )}}
\\ \\
\displaystyle{\phantom{aa}=\,
\kappa^m(x)
\frac{\sigma^{m}(\lambda , x)
\hat \tau^{m+1}(\lambda , x+\eta )}{\sigma^{m}(\lambda , x+\eta )
\sigma^{m+1}(\lambda , x)}
-\kappa^m(x)
\frac{\sigma^{m+1}(\lambda , x+\eta )
\hat \tau^{m}(\lambda , x)}{\sigma^{m}(\lambda , x+\eta )
\sigma^{m+1}(\lambda , x)}},
\end{array}
\eeq
where
\beq\label{f7}
\kappa^m(x)=\kappa \, \frac{\rho^m(x)\rho^{m+1}(x+\eta )}{\rho^m(x+\eta )
\rho^{m+1}(x)}.
\eeq
The further calculation is similar to the one performed in the
previous section. 
The both sides of $(\ref{f6})$ 
are elliptic functions of $\lambda$ with 
simple poles at $\lambda =\lambda_i^{m+1}(x)$ 
and $\lambda =\lambda_i^m(x-\eta )$.
Equating the residues at these poles 
and making the shifts
$m\to m-1$, $x\to x-\eta$ when necessary, we get the equations
\beq\label{f8}
\begin{array}{l}
\sigma^{m-1}(\lambda_i^{m}(x), x+\eta )\hat \tau^{m}(\lambda_i^m(x), x)
=
\kappa^{m-1}(x)\sigma^{m-1}(\lambda_i^{m}(x), x)
\hat \tau^{m}(\lambda_i^m(x), x+\eta )
\\ \\
\phantom{aaaaaaaaaaaaaaaaa}-
\kappa^{m-1}(x)\sigma^{m}(\lambda_i^{m}(x), x+\eta )
\hat \tau^{m-1}(\lambda_i^m(x), x),
\end{array}
\eeq
\beq\label{f8a}
\begin{array}{l}
\sigma^{m+1}(\lambda_i^{m}(x), x-\eta )\hat \tau^{m}(\lambda_i^m(x), x)
=
\kappa^{m}(x-\eta )\sigma^{m+1}(\lambda_i^{m}(x), x)
\hat \tau^{m}(\lambda_i^m(x), x-\eta )
\\ \\
\phantom{aaaaaaaaaaaaaaaaa}-
\kappa^{m}(x-\eta )\sigma^{m}(\lambda_i^{m}(x), x-\eta )
\hat \tau^{m+1}(\lambda_i^m(x), x).
\end{array}
\eeq
Substituting $\lambda =\lambda_i^m(x)$ and $\lambda 
=\lambda_i^{m+1}(x\! -\! \eta )$ into (\ref{f6}),
we see that one term vanishes. 
Making the shifts $m\to m-1$, $x\to x-\eta$ when necessary,
we get the equations
\beq\label{f9}
\begin{array}{l}
\kappa^{m}(x)\sigma^{m+1}(\lambda_i^{m}(x), x+\eta )
\hat \tau^{m}(\lambda^m_i(x), x)
=\sigma^{m+1}(\lambda_i^{m}(x), x)
\hat \tau^{m}(\lambda_i^m(x), x+\eta )
\\ \\
\phantom{aaaaaaaaaaaaaaaaa}
-\sigma^{m}(\lambda_i^{m}(x), x+\eta )
\hat \tau^{m+1}(\lambda_i^m(x), x),
\end{array}
\eeq
\beq\label{f9a}
\begin{array}{l}
\kappa^{m-1}(x-\eta )\sigma^{m-1}(\lambda_i^{m}(x), x-\eta )
\hat \tau^{m}(\lambda^m_i(x), x)
=\sigma^{m-1}(\lambda_i^{m}(x), x)
\hat \tau^{m}(\lambda_i^m(x), x-\eta )
\\ \\
\phantom{aaaaaaaaaaaaaaaaa}
-\sigma^{m}(\lambda_i^{m}(x), x-\eta )
\hat \tau^{m-1}(\lambda_i^m(x), x).
\end{array}
\eeq
Consider equations (\ref{f8}) and (\ref{f9}). They can be regarded as a
system of linear
equations for $\hat \tau^m(\lambda_i^m(x),x)$ and 
$\hat \tau^m(\lambda_i^m(x), x\! +\! \eta )$. In their turn,
equations (\ref{f8a}) and (\ref{f9a}) can be regarded as a
system of linear
equations for $\hat \tau^m(\lambda_i^m(x),x)$ and 
$\hat \tau^m(\lambda_i^m(x), x\! -\! \eta )$. Solving them for
$\hat \tau^m(\lambda_i^m(x),x)$ with the help of the Cramer's rule,
as in the previous section, and equating the results, we get the equations
\beq\label{f10}
\begin{array}{c}
(\kappa^{m-1}(x))^{-1}\sigma^{m+1}(\lambda_i^m(x), x)
\sigma^{m}(\lambda_i^m(x), x-\eta )\sigma^{m-1}(\lambda_i^m(x), x+\eta )
\\ \\
+(\kappa^{m}(x-\eta ))^{-1}\sigma^{m+1}(\lambda_i^m(x), x-\eta )
\sigma^{m}(\lambda_i^m(x), x+\eta )\sigma^{m-1}(\lambda_i^m(x), x)
\\ \\
=\, 
\kappa^{m-1}(x-\eta )\sigma^{m+1}(\lambda_i^m(x), x)
\sigma^{m}(\lambda_i^m(x), x+\eta )\sigma^{m-1}(\lambda_i^m(x), x-\eta )
\\ \\
+\kappa^{m}(x)\sigma^{m+1}(\lambda_i^m(x), x+\eta )
\sigma^{m}(\lambda_i^m(x), x-\eta )\sigma^{m-1}(\lambda_i^m(x), x).
\end{array}
\eeq
These are equations of motion for the field analogue of the deformed 
RS model on the space-time lattice. They resemble equations 
(\ref{int5}) and reduce to them if one sets
$\lambda_i^m(x)=x_i^m +x$ and $\kappa^m(x)=\kappa =\mbox{const}$.
In the limit $\kappa \to 0$, when the right hand side vanishes,
equations (\ref{f10}) reduce to the fully discrete version of the 
field extension of the RS model \cite{ZZ21} (see also \cite{DNY15},
where similar equations were obtained from the elliptic Lax pair
of general form).
Note also that equations (\ref{f10})
can be written entirely in terms of the function 
$\tau^m(\lambda , x)$ (\ref{f3}):
\beq\label{f11}
\begin{array}{c}
\tau^{m+1}(\lambda_i^m(x), x)
\tau^{m}(\lambda_i^m(x), x-\eta )\tau^{m-1}(\lambda_i^m(x), x+\eta )
\\ \\
+\tau^{m+1}(\lambda_i^m(x), x-\eta )
\tau^{m}(\lambda_i^m(x), x+\eta )\tau^{m-1}(\lambda_i^m(x), x)
\\ \\
=\, 
\tau^{m+1}(\lambda_i^m(x), x)
\tau^{m}(\lambda_i^m(x), x+\eta )\tau^{m-1}(\lambda_i^m(x), x-\eta )
\\ \\
+\tau^{m+1}(\lambda_i^m(x), x+\eta )
\tau^{m}(\lambda_i^m(x), x-\eta )\tau^{m-1}(\lambda_i^m(x), x).
\end{array}
\eeq
In the next subsection we will discuss continuous time limits
of these equations.

\subsection{Continuous time limits}

For the continuous time limit, we rewrite equations (\ref{f10})
in the form
\beq\label{ct1}
\begin{array}{c}
\displaystyle{
\frac{\kappa^m (x\! -\! \eta )}{\kappa^{m-1}(x)}\prod_j
\frac{\sigma (\lambda_i^m(x)-\lambda_j^{m+1}(x))
\sigma (\lambda_i^m(x)-\lambda_j^{m}(x\! -\! \eta ))
\sigma (\lambda_i^m(x)-
\lambda_j^{m-1}(x\! +\! \eta ))}{\sigma (\lambda_i^m(x)-\lambda_j^{m-1}(x))
\sigma (\lambda_i^m(x)-\lambda_j^{m+1}(x\! -\! \eta ))
\sigma (\lambda_i^m(x)-\lambda_j^{m}(x\! +\! \eta ))}}
\\ \\
\displaystyle{
=\, -1+\kappa^m(x)\kappa^m (x\! -\! \eta )\prod_j
\frac{\sigma (\lambda_i^m(x)-\lambda_j^{m+1}(x\! +\! \eta ))
\sigma (\lambda_i^m(x)-
\lambda_j^{m}(x\! -\! \eta ))}{\sigma (\lambda_i^m(x)-
\lambda_j^{m}(x\! +\! \eta ))
\sigma (\lambda_i^m(x)-\lambda_j^{m+1}(x\! -\! \eta ))}}
\\ \\
\displaystyle{
+\, \kappa^{m-1}(x\! -\! \eta )\kappa^m (x\! -\! \eta )\prod_j
\frac{\sigma (\lambda_i^m(x)-\lambda_j^{m+1}(x))
\sigma (\lambda_i^m(x)-
\lambda_j^{m-1}(x\! -\! \eta ))}{\sigma (\lambda_i^m(x)-
\lambda_j^{m-1}(x))
\sigma (\lambda_i^m(x)-\lambda_j^{m+1}(x\! -\! \eta ))}}
\end{array}
\eeq
and assume that the $\lambda_j^m$'s and $\rho^m$ behave smoothly 
when the time changes, i.e.
$$
\begin{array}{l}
\lambda_i^{m\pm 1}(x)=\lambda_i(x)\pm \varepsilon \dot \lambda_i(x)+
\frac{1}{2}\, \varepsilon^2 \ddot \lambda_i(x)+O(\varepsilon^3),
\\ \\
\rho^{m\pm 1}(x)=\rho (x)\pm \varepsilon \dot \rho (x)+O(\varepsilon^2).
\end{array}
$$
The limit is straightforward. 
We should expand the equation in powers of 
$\varepsilon$ as $\varepsilon \to 0$.
If $\kappa^2 \neq 1$, then the first non-vanishing order is
$\varepsilon$, and we get the equations
\beq\label{ct2}
\begin{array}{c}
\displaystyle{
\ddot \lambda_i(x)+\sum_j \Bigl [\dot \lambda_i(x)
\dot \lambda_j(x\! -\! \eta ) \zeta \Bigl (\lambda_i(x)\! -\! 
\lambda_j(x\! -\! \eta )\Bigr ) \! +\! 
\dot \lambda_i(x)
\dot \lambda_j(x\! +\! \eta ) \zeta \Bigl (\lambda_i(x) \! -\! 
\lambda_j(x\! +\! \eta )\Bigr )\Bigr ]}
\\ \\
\displaystyle{
-\, 2 \sum_{j\neq i}\dot \lambda_i(x)
\dot \lambda_j(x) \zeta \Bigl (\lambda_i(x)-\lambda_j(x)\Bigr )+
\Bigl (c(x-\eta )-c(x)\Bigr )\dot \lambda_i(x)=0,}
\end{array}
\eeq
where
$$
c(x)=\frac{\dot \rho (x+\eta )}{\rho (x+\eta )}-
\frac{\dot \rho (x)}{\rho (x)}.
$$
Summing the equations (\ref{ct2}) over all $i$ and taking into account
the condition (\ref{f2}), we find that $c(x)$ is expressed in terms
of the $\lambda_i$'s as follows:
\beq\label{ct3}
c(x)=\Bigl (\sum_k \dot \lambda_k(x)\Bigr )^{-1}
\sum_{i,j} \dot \lambda_i(x)
\dot \lambda_j(x\! +\! \eta ) \zeta \Bigl (\lambda_i(x) \! -\! 
\lambda_j(x\! +\! \eta )\Bigr ).
\eeq
Equations (\ref{ct2}), (\ref{ct3}) were obtained in \cite{ZZ21}
as equations of motion for the field analogue of the RS model.
If $\kappa^2=1$, then the order
$\varepsilon$ gives the identity $0=0$, and one 
has to expand up to the order 
$\varepsilon^2$. In this way it is possible to obtain the field
analogue of equations (\ref{d4}). 

Like in the previous section, another continuum limit is possible.
We introduce the lattice fields $\varphi_i^m(x)$, $\omega^m(x)$ by setting
\beq\label{ct4}
\lambda_i^m(x)=\varphi_i^m(x+m\eta ), \quad
\rho^m(x)=\omega^m (x+m\eta )
\eeq
and assume that these new fields have a smooth continuous time
limit in the sense that
$$
\begin{array}{l}
\varphi_i^{m\pm 1}(x)=\varphi_i(x)\pm \varepsilon \dot \varphi_i(x)+
\frac{1}{2}\, \varepsilon^2 \ddot \varphi_i(x)+O(\varepsilon^3),
\\ \\
\omega^{m\pm 1}(x)=\omega (x)\pm \varepsilon \dot 
\omega (x)+O(\varepsilon^2).
\end{array}
$$
In terms of the new fields, equations (\ref{ct1}) read:
\beq\label{ct5}
\begin{array}{c}
\displaystyle{
\frac{\kappa^m (x\! -\! \eta )}{\kappa^{m-1}(x)}\prod_j
\frac{\sigma (\varphi_i^m(x)-\varphi_j^{m+1}(x\! +\! \eta))
\sigma (\varphi_i^m(x)-\varphi_j^{m}(x\! -\! \eta ))
\sigma (\varphi_i^m(x)-
\varphi_j^{m-1}(x))}{\sigma (\varphi_i^m(x)-\varphi_j^{m}(x\! +\! \eta))
\sigma (\varphi_i^m(x)-\varphi_j^{m-1}(x\! -\! \eta ))
\sigma (\varphi_i^m(x)-\varphi_j^{m+1}(x))}}
\\ \\
\displaystyle{
=\, -1+\kappa^m(x)\kappa^m (x\! -\! \eta )\prod_j
\frac{\sigma (\varphi_i^m(x)-\varphi_j^{m+1}(x\! +\! 2\eta ))
\sigma (\varphi_i^m(x)-
\varphi_j^{m}(x\! -\! \eta ))}{\sigma (\varphi_i^m(x)-
\varphi_j^{m}(x\! +\! \eta ))
\sigma (\varphi_i^m(x)-\varphi_j^{m+1}(x))}}
\\ \\
\displaystyle{
+\, \kappa^{m-1}(x\! -\! \eta )\kappa^m (x\! -\! \eta )\prod_j
\frac{\sigma (\varphi_i^m(x)-\varphi_j^{m-1}(x\! -\! 2\eta))
\sigma (\varphi_i^m(x)-
\varphi_j^{m+1}(x\! +\! \eta ))}{\sigma (\varphi_i^m(x)-
\varphi_j^{m-1}(x\! -\! \eta ))
\sigma (\varphi_i^m(x)-\varphi_j^{m+1}(x))}.}
\end{array}
\eeq
The limit $\varepsilon \to 0$ is consistent if $\kappa =\varepsilon$. 
Then in the order $\varepsilon$ we obtain the equations
\beq\label{ct6}
\begin{array}{c}
\displaystyle{
\ddot \varphi_i(x)+\sum_j \Bigl [\dot \varphi_i(x)
\dot \varphi_j(x\! -\! \eta ) \zeta \Bigl (\varphi_i(x)\! -\! 
\varphi_j(x\! -\! \eta )\Bigr ) \! +\! 
\dot \varphi_i(x)
\dot \varphi_j(x\! +\! \eta ) \zeta \Bigl (\varphi_i(x) \! -\! 
\varphi_j(x\! +\! \eta )\Bigr )\Bigr ]}
\\ \\
\displaystyle{
-\, 2 \sum_{j\neq i}\dot \varphi_i(x)
\dot \varphi_j(x) \zeta \Bigl (\varphi_i(x)-\varphi_j(x)\Bigr )
-\p_t \log w(x) \dot \varphi_i(x)}
\\ \\
\displaystyle{+\, w(x)w(x+\eta )G_i^+ +w(x)w(x-\eta )G_i^-=0},
\end{array}
\eeq
where
\beq\label{ct7}
w(x)=\frac{\omega (x+\eta )\omega (x-\eta )}{\omega^2(x)},
\eeq
\beq\label{ct8}
\begin{array}{c}
\displaystyle{
G_i^+=\frac{\sigma (\varphi_i(x)\! -\! \varphi_i (x\! +\! 2\eta ))
\sigma (\varphi_i(x)\! -\! 
\varphi_i (x\! -\! \eta ))}{\sigma (\varphi_i(x)\! -\! 
\varphi_i (x\! +\! \eta ))}}
\\ \\
\displaystyle{\phantom{aaaaaaaaaaaaaaaaaaaa}
\times \, \prod_{j\neq i}
\frac{\sigma (\varphi_i(x)\! -\! \varphi_j (x\! +\! 2\eta ))
\sigma (\varphi_i(x)\! -\! 
\varphi_j (x\! -\! \eta ))}{\sigma (\varphi_i(x)\! -\! 
\varphi_j (x\! +\! \eta ))\sigma (\varphi_i(x)\! -\! 
\varphi_j (x))}}
\end{array}
\eeq
and $G_i^-$ differs from $G_i^+$ by the change $\eta \to -\eta$. 
Summing equations (\ref{ct6}) over all $i$, we obtain the equation
\beq\label{ct9}
\p_t \log w(x) \sum_i \dot \varphi_i(x)=F(x)+
w(x)w(x+\eta )\sum_i G_i^+ +w(x)w(x-\eta )\sum_i G_i^-,
\eeq
where
\beq\label{ct10}
F(x)=\sum_{i,j}\Bigl [
\dot \varphi_i (x)\dot \varphi_j (x\! +\! \eta )\zeta \Bigl (
\varphi_i (x)-\varphi_j (x\! +\! \eta )\Bigr )+
\dot \varphi_i (x)\dot \varphi_j (x\! -\! \eta )\zeta \Bigl (
\varphi_i (x)-\varphi_j (x\! -\! \eta )\Bigr )\Bigr ].
\eeq
Equations (\ref{ct6}) together with (\ref{ct9}) form a system 
of $N+1$ differential equations for the $N+1$ fields 
$\varphi_j (x)=\varphi_j (x,t)$ ($j=1, \ldots , N$), $w(x)=w(x,t)$.
In contrast to equations (\ref{ct2}), where the extra field was not
dynamical and could be excluded from the equations of motion, in the 
present case the extra field $w$ is dynamical. Equations
(\ref{ct6}), (\ref{ct9}) provide the field extension of the 
deformed RS system (\ref{int3}). 

\section{Concluding remarks}

In this paper we have obtained integrable time discretization of
the deformed RS system. The method is based on the explicit form of the
B\"acklund transformation which is obtained as equations connecting 
the dynamics of poles and zeros of double-Bloch solutions $\psi$ to the 
linear problem for the Toda lattice with constraint of type B. 
As in other systems of the Calogero-Moser 
and Ruijsenaars-Schneider type, 
the B\"acklund transformation is the passage from poles to zeros
of the $\psi$-function, and this is interpreted as one step forward
in the discrete time evolution. Possible continuum limits of the 
discrete time equations obtained were also discussed. One of them
gives equations of motion of the deformed RS system while another one 
gives equations of motion for poles of elliptic solutions to the 
semi-discrete BKP equation \cite{RZ20}. 

We have also shown that
the discrete time equations of motion for 
the deformed RS system describe evolution of poles of elliptic
solutions to the fully discrete BKP equation.
Besides, by considering more general elliptic solutions of the latter
(the so-called elliptic families), we have obtained a field extension 
of the deformed RS model on a space-time lattice.

It is an important open question whether the obtained equations
admit any commutation representation, i.e. whether they can be represented
as a relation between matrices whose matrix elements depend on the 
dynamical variables, like (discrete versions of) Lax and Manakov triple 
representations or representation of the Zakharov-Shabat 
type. We hope to address this problem elsewhere.

Finally, we would like to mention 
that the discrete time equations
of motion for the RS system (\ref{int1}) 
mysteriously coincide with the nested
Bethe ansatz equations arising in the theory of quantum integrable
systems with elliptic $R$-matrix. It is still not clear whether 
this is just a coincidence or this fact has some profound reasons. 
In this connection it is natural to ask whether the 
discrete time equations (\ref{int5}) for the deformed RS system 
have any relation to quantum integrable systems. To wit, 
the question is whether there
exists any quantum integrable system solved by Bethe ansatz
or any other method whose Bethe-like equations 
would be of the form (\ref{int5}).

\section*{Appendix A: The Weierstrass functions}
\addcontentsline{toc}{section}{Appendix A: The Weierstrass functions}
\def\theequation{A\arabic{equation}}
\def\theHequation{\theequation}
\setcounter{equation}{0}

In this appendix we present the definition and main properties of the 
Weierstrass functions: 
the $\sigma$-function, the $\zeta$-function and the $\wp$-function
which are used in the main text.

Let $\omega$, $\omega '$ be complex numbers such that 
${\rm Im} (\omega '/ \omega )>0$.
The Weierstrass $\sigma$-function 
with quasi-periods $2\omega$, $2\omega '$ 
is defined by the following infinite product over the lattice
$2\omega m+2\omega ' m'$, $m,m'\in \ZZ$:
\beq\label{A1}
\sigma (x)=\sigma (x |\, \omega , \omega ')=
x\prod_{s\neq 0}\Bigl (1-\frac{x}{s}\Bigr )\, 
e^{\frac{x}{s}+\frac{x^2}{2s^2}},
\quad s=2\omega m+2\omega ' m' \quad m, m'\in \ZZ .
\eeq 
It is an odd quasiperiodic function with two linearly independent
quasi-periods in the complex plane. 
The expansion around $x=0$ is
\beq\label{A1a}
\sigma (x)=x+O(x^5), \quad x\to 0.
\eeq
The monodromy properties of the $\sigma$-function 
under shifts by the quasi-periods
are as follows:
\beq\label{A4}
\begin{array}{l}
\sigma (x+2\omega )=-e^{2\zeta (\omega )(x+\omega )}\sigma (x),
\\ \\
\sigma (x+2\omega ' )=-e^{2\zeta (\omega ')(x+\omega ' )}\sigma (x).
\end{array}
\eeq
Here $\zeta (x)$ is the
Weierstrass $\zeta$-function defined as
\beq\label{A4a}
\zeta (x)=\frac{\sigma '(x)}{\sigma (x)}.
\eeq
As $x\to 0$,
\beq\label{A2a}
\zeta (x)=\frac{1}{x} +O(x^3), \quad x\to 0.
\eeq

The Weierstrass $\wp$-function is defined as
$\wp (x)=-\zeta '(x)$. 
It is an even double-periodic function with periods $2\omega , 2\omega '$
and with second order poles at the points 
of the lattice $s=2\omega m+2\omega ' m'$ with integer $m, m'$.
As $x\to 0$, $\wp (x)=x^{-2}+O(x^2)$.

The Weierstrass functions obey many non-trivial identities.
Here we present the two which are necessary for the 
calculations leading to equation (\ref{d4}):
\beq\label{A5}
\zeta (x+\eta )+\zeta (x-\eta )-2\zeta (x)=
\frac{\wp '(x)}{\wp (x)-\wp (\eta )},
\eeq
\beq\label{A6}
\wp (x+\eta )-\wp (x-\eta )=-\, \frac{\wp '(x)\wp '(\eta )}{(\wp (x)-
\wp (\eta ))^2}.
\eeq
The proof is standard. The both sides are elliptic functions of $x$,
and the singular terms in the both sides coincide. Therefore, the difference
between the left and right hand sides is a constant which can be found
by putting $x$ to some special value.

\section*{Acknowledgments}

\addcontentsline{toc}{section}{Acknowledgments}

This work has been supported in part within the framework of the
HSE University Basic Research Program.

\end{document}